\documentstyle[aps,preprint,psfig]{revtex}

\newcommand{\be}{\begin{equation}}
\newcommand{\ee}{\end{equation}}
\newcommand{\Dlt}{\Delta}
\newcommand{\dlt}{\delta}
\newcommand{\prt}{\partial}
\newcommand{\br}{{\bf x}}
\newcommand{\bp}{{\bf p}}
\newcommand{\bq}{{\bf q}}
\newcommand{\bk}{{\bf k}}
\newcommand{\bt}{\beta}

\newcommand{\ep}{\varepsilon}
\newcommand{\al}{\alpha}
\newcommand{\ra}{\rightarrow}
\newcommand{\sgm}{\sigma}

\newcommand{\om}{\xi\hspace{.2pt}}
\newcommand{\Om}{\Omega}

\newcommand{\dgr}{\dagger}

\tightenlines

\begin{document}

\draft

\title{Gapless Hartree-Fock-Bogoliubov Approximation for Bose Gases}
\author{V.I. Yukalov$^{1,2}$ and H. Kleinert$^1$}

\address{$^1$Institut f\"ur Theoretische Physik, \\
Freie Universit\"at Berlin, Arnimallee 14, D-14195 Berlin, Germany}
\address{$^2$Bogoliubov Laboratory of Theoretical Physics, \\
Joint Institute for Nuclear Research, Dubna 141980, Russia}

\maketitle

\vskip 2cm

\begin{abstract}

A dilute Bose system with Bose-Einstein condensate is considered.
It is shown that the Hartree-Fock-Bogolubov approximation can be made both
conserving as well as gapless. This is achieved by taking into account all
physical normalization conditions, that is, the normalization condition
for the condensed particles and that for the total number of particles.
Two Lagrange multipliers, introduced for preserving these normalization
conditions, make the consideration completely self-consistent.

\end{abstract}

\vskip 1.5cm


\newpage

\section{Introduction}

The physical properties of Bose gases exhibiting Bose-Einstein condensation
are now a topic of intensive research, both experimental and theoretical
[1--5]. The physics of dilute weakly interacting Bose gas has been studied
especially well using Bogoliubov's model and operator techniques
\cite{6,7,8,9}. Despite the apparent simplicity of the system, some
principal problems in its theoretical description have not yet been well
understood. Most notorious is the so-called dilemma of conserving versus
gapless approximations, for which we shall offer a possible solution in this
paper.

The above dilemma arises, when one attempts to describe the weakly
interacting Bose gas at finite (not asymptotically small) diluteness
parameter. It is absent in the original Bogoliubov approximation [6--9],
which is valid only at very low temperatures $T\ra 0$ and asymptotically
weak interaction. The first natural extension to finite temperatures and
diluteness parameters is a bosonic version of the Hartree-Fock-Bogoliubov
(HFB) approximation. This is a self-consistent approximation which is
guaranteed to respect all conservation laws that follow from the underlying
symmetries of the Hamiltonian. However, this approximation turns out to
render a gap in the spectrum of collective excitations in the condensed
phase in which the global U(1) gauge symmetry of the theory is broken.
Also the Girardeau-Arnowitt approximation \cite{10} displays this gap, since
it is  equivalent to the HFB approximation. This contradicts the fundamental
theorems of Hugenholtz and Pines \cite{11}, Goldstone \cite{12}, and
Bogoliubov [9], according to which the spectrum of collective excitations
in the symmetry-broken phase has to be gapless. It is well known
\cite{13,14,14a} that the HFB approximation can be formulated as a variational
approximation. There exist several other variational approximations, the
so-called $\Phi$-derivable approximations or higher effective actions
\cite{15,16,17,18,18a}, which also respect the conservation laws but lead
to a gap in the spectrum. When trying to remove the gap, one usually
violates the conservation laws and runs into other thermodynamic
inconsistencies. The various approximations that have been studied are
typically classified as either conserving or gapless. This classification
and the related dilemma were first emphasized by
Hohenberg and Martin
\cite{19}, and later discussed in many publications, for instance, by Baym
and Grinstein \cite{18}. A very detailed discussion of this problem, with
many citations, has recently been done by Andersen [4].

In order to remove the gap in the HFB approximation, one often invokes
a trick of neglecting the anomalous averages, calling this the "Popov
approximation". However, a glance at the original works by Popov
\cite{20,21,22,23} shows that he has never suggested this trick. What he
actually considered was a narrow region of temperatures $T$ in the vicinity
of the condensation temperature $T_c$. When $T\ra T_c$, then the condensate
density tends to zero. The anomalous averages, being proportional to the
condensate density, tend to zero together with the latter, when $T\ra T_c$.
As a result, their contribution becomes automatically small, without any
special assumptions. Far below $T_c$, however, the anomalous averages can be
very large, and Popov has never proposed to neglect them there. It is
straightforward to demonstrate by direct calculations that at low temperatures
$T\ll T_c$ the anomalous averages become of the same order as the normal
averages. They can even be much larger than the latter \cite{24}, so that
Popov would certainly not have been keen on proposing the so-called "Popov
approximation".

Moreover, preserving the anomalous averages for the phase with broken U(1)
gauge symmetry is principally important, since their negligence makes the
system unstable \cite{24,25}.

There have been several attempts to make the HFB approximation gapless by
adding to this approximation some higher-order terms involving the
Bethe-Salpeter or $T$-matrix approximations. The idea to modify in this way
the mean-field HFB approximation goes back to Kirzhnits and Linde \cite{26}
and Baym and Grinstein \cite{18}. Several such modifications of the HFB
approximation have been considered, in which additional terms are either
motivated by higher-order approximations \cite{27,28} or just added
phenomenologically \cite{29,30}. This type of modifications
\cite{26,27,28,29,30} possesses a number of deficiencies, which do not
permit to accept this approach as a solution of the problem. A good
analysis of these deficiencies has already been done by Baym and Grinstein
\cite{18} and recently by Andersen [4].

First of all, the method of adding to the theory some additional terms in
order to cancel the spectrum gap is ambiguous, not following from a general
physical law. As a result, the additional terms are not uniquely defined.
There are no general grounds to decide which of the variants is better.

Second, the way of mixing different approximations is not self-consistent.
This is what Bogoliubov [9] called the "mismatch of approximations".
Although such a mismatch can make the gap disappear, it is usually
inconsistent with some thermodynamic equations. For instance, the chemical
potential defined by the Hugenholtz-Pines theorem to yield a gapless
spectrum, does not coincide with the chemical potential found from the
minimization of the thermodynamic potential with respect to the condensate
density. This discrepancy is a general feature of all non-self-consistent
approximations, due to which they cannot properly be called gapless [4].

Moreover, if the price of making the spectrum gapless is that the
thermodynamic potential cannot be minimized, this implies that the system
becomes thermodynamically unstable. Thus one has the unpleasant alternative:
either the system is gapless but unstable or seemingly stable but with a
gap. This is a particular case of the general problem of thermodynamic
self-consistency  \cite{31,32}.

Problems with the thermodynamics of the model have the unpleasant
consequences of modifying the order of the phase transition \cite{31,32}.
This happens for all approaches with mismatched approximations. Baym and
Grinstein \cite{18} emphasized that  attempts to modify the HFB approximation
in a non-self-consistent way lead to a first-order condensation transition,
instead of the observed second-order one. Further works [4,31,32] confirmed
that it is a common feature of all non-self-consistent mean-field
approximations.

In the present paper, we show that the HFB approximation can be made both
conserving {\em and\/} gapless, while avoiding the mismatch of approximations,
thus being completely self-consistent. The solution of the problem is possible
by taking into account two existing normalization conditions instead of one
in all previous approximations. This makes the HFB approximation self-consistent
and gapless, without any tricks or additional terms. As a consequence,
condensation transition becomes second order, as it should be.

Throughout the paper, the system of units is employed with  Planck and
Boltzmann constants equal to unity, $\hbar\equiv 1$, $k_B\equiv 1$.

\section{Grand-Canonical Hamiltonian}

We consider a dilute Bose gas, whose particle interactions are modeled by
the contact potential
\be
\label{1}
V(\br) =g\dlt(\br) \; ,
\ee
with the interaction intensity
\be
\label{2}
g \equiv 4\pi \; \frac{a_s}{m}
\ee
expressed through the $s$-wave scattering length $a_s$ and particle mass
$m$. The energy operator reads
\be
\label{3}
\hat H = \int d^3x\, \psi^\dgr(\br) \left ( -\; \frac{\nabla^2}{2m} \right )
\psi(\br)\;  + \frac{g}{2}\, \int d^3x  \,
\psi^\dgr(\br) \psi^\dgr(\br) \psi(\br)\psi(\br) ,\;
\ee
with the field operators $\psi(\br)$ satisfying the Bose commutation
relations. The operator counting the total number of particles is
\be
\label{4}
\hat N = \int d^3x\,\psi^\dgr(\br)\psi(\br) .
\ee
In what follows, we treat a uniform system in thermodynamic equilibrium,
since this is the simplest situation where the above-discussed problems
reveal themselves.

If the temperature in a Bose system falls below the condensation temperature
$T_c$, U(1) gauge symmetry becomes broken. The symmetry breaking is taken
into account by the Bogoliubov shift [8,9] of the field operator
\be
\label{5}
\psi(\br) \; \longrightarrow \; \Psi +\psi_1(\br) \; ,
\ee
where $\Psi$ is the condensate {\em order parameter\/}, which in uniform
systems is independent of $\br$, and $\psi_1(\br)$ is the field operator of
the uncondensed particles, satisfying the same Bose commutation relations as
$\psi(\br)$. Another method of first separating the zero-momentum components
of the field and then replacing them by classical numbers [6,7] is completely
equivalent to the shift (5), as has been rigorously proved by Ginibre \cite{33}.
The field $\psi_1(\br)$ has no zero-momentum component so that
\be
\label{6}
\langle\psi_1(\br)\rangle \; = \; 0
\ee
and $\Psi$ and $\psi_1(\br)$ are orthogonal to each other:
\be
\label{7}
\int \Psi^* \psi_1(\br)\; d^3x = 0.
\ee
The condensate order parameter $\Psi$ defines the density of condensed
particles
\be
\label{8}
\rho_0 = |\Psi|^2 \; .
\ee
Above $T_c$ where $\Psi$ vanishes, there is no condensate. Below $T_c$, one
has $\Psi\neq 0$ and thus a finite condensate density $\rho_0$.

The free energy of the system is
\be
\label{9}
F = - T  \ln {\rm Tr} \; e^{-\bt\hat H} \; ,
\ee
with $\bt\equiv 1/T$. After substituting shift (5) into $\hat H$, the model
contains two field variables, the condensate order parameter $\Psi$ and the 
space-dependent field $\psi_1(\br)$. They give rise to {\em two\/} 
normalization conditions. One is related to the definition of the total 
number of particles
\be
\label{10}
N \; = \;\langle\hat N\rangle \; .
\ee
A second normalization condition fixes the number of condensed particles
\be
\label{11}
N_0 = \rho_0 V = |\Psi|^2 V \; .
\ee

In stable equilibrium, the free energy gains a minimum under the normalization
conditions (10) and (11). This conditional minimization is equivalent to the
unconditional minimum of the grand-canonical  potential
\be
\label{12}
\Om = - T \ln {\rm Tr}\; e^{-\bt H} \; ,
\ee
with the grand-canonical  Hamiltonian
\be
\label{13}
H = \hat H - \mu_0 N_0 - \mu\hat N \; ,
\ee
in which $\mu_0$ and $\mu$ are the Lagrange multipliers enforcing
 the normalization conditions (10) and (11). The minimum of the
grand-canonical  potential (12) is determined by the equations
\be
\label{14}
\frac{\prt\Om}{\prt N_0} = 0 \; , \qquad
\frac{\prt^2\Om}{\prt N_0^2} > 0 \; .
\ee
The first derivative is given by the expectation value
\be
\label{15}
\frac{\prt\Om}{\prt N_0}  = \; \left\langle  \frac{\prt H}{\prt N_0}
\right\rangle \; .
\ee
The second derivative is calculated from
\be
\label{16}
\frac{\prt^2\Om}{\prt N_0^2} = \; \left\langle \frac{\prt^2 H}{\prt N^2_0}
\right\rangle  + \;
\bt\; \Dlt^2\left ( \frac{\prt H}{\prt N_0} \right ) \; ,
\ee
with the notation
$$
\Dlt^2(\hat {\cal O})\equiv\langle\hat {\cal O}^2\rangle-\langle
\hat {\cal O}\rangle^2
$$
for the dispersion of an operator ${\hat {\cal O}}$.

For a uniform system, the field operator of the uncondensed particles is
expandable in plane waves as
\be
\label{17}
\psi_1(\br) = \sum_{\bk\neq 0} a_\bk  \;
\frac{e^{i\bk\cdot\br}}{\sqrt{V}} \; .
\ee
Performing the Bogoliubov shift (5), together with the expansion (17),
we obtain for the grand-canonical  Hamiltonian (13) the  sum of five terms
\be
\label{18}
H = \sum_{n=0}^4 H^{(n)} \; ,
\ee
depending on the
the power of the operators $\psi_1(\br)$.
The zero-order term
\be
\label{19}
H^{(0)} = \left ( \frac{1}{2}\; \rho_0 g - \ep\right ) N_0 \; ,
\ee
with
\be
\label{20}
\ep \equiv \mu_0 + \mu \; ,
\ee
is free of $\psi_1(\br)$. The first-order term in Eq. (19) vanishes, since
the decomposition (17) contains only nonzero momenta. The second-order term is
\be
\label{21}
H^{(2)} = \sum_{\bk\neq 0} \left ( \frac{k^2}{2m} +  2\rho_0g - \mu
\right ) a_\bk ^\dgr a_\bk  + \frac{\rho_0 g
}{2}\; \sum_{\bk\neq 0} \left ( a_\bk ^\dgr a_{-\bk} ^\dgr + a_{-\bk}  a_\bk
\right ) \; .
\ee
In the third-order term
\be
\label{22}
H^{(3)} = \sqrt{\frac{\rho_0}{V}} \;g\, {\sum_{\bp,\bq}}'  \left (
a_\bq ^\dgr a_{\bp+\bq} a_{-\bp } + a_{-\bp }^\dgr a_{\bp+\bq}^\dgr a_\bq
\right ) \; ,
\ee
the prime on the summation symbol implies that $\bp\neq 0$, $\bq\neq 0$, and
$\bp+\bq\neq 0$. In the fourth-order term
\be
\label{23}
H^{(4)}  = \frac{g}{2V} \; \sum_\bk   {\sum_{\bp,\bq}}'
a_\bp^\dgr a_\bq ^\dgr a_{\bp+\bk} a_{\bq-\bk} \; ,
\ee
the summation does not include any zero-momentum operators, so that the prime
means that $\bp\neq 0$, $\bq\neq 0$, $\bk+\bp\neq 0$, and $\bk-\bq\neq 0$.

The field operators in momentum space $a_\bk $ satisfy the following
conditions. From Eq. (7) we have
\be
\label{24}
\langle a_\bk \rangle  \; = \; 0 \qquad (\bk\neq 0) \; ,
\ee
and owing to the uniformity of the system:
\be
\label{25}
\langle a_\bk ^\dgr a_\bp\rangle  \; =
\dlt_{\bk,\bp}\langle a_\bk ^\dgr a_\bk \rangle  \; , \qquad
\langle a_\bk  a_\bp\rangle  \; =
\dlt_{-\bk,\bp} \langle  a_\bk  a_{-\bk}  \rangle  \; .
\ee

Note that $\mu$ in Eq. (21) is the chemical potential enforcing the 
normalization condition (10). In a system without Bose-Einstein condensate, 
there is no need to introduce another Lagrange multiplier. However, as soon 
as the gauge symmetry is broken and a Bose-Einstein condensate appears, the 
theory acquires the new variable $\Psi$, the order parameter of the condensate, 
which satisfies the normalization condition (11). For the self-consistency 
of the theory, it is then necessary to take account of this additional 
normalization condition, which requires the second Lagrange multiplier $\mu_0$. 
Without the latter, the theory cannot be made self-consistent, and the 
normalization condition for the condensed particles cannot be guaranteed.

It is worth emphasizing that $\mu$ is the chemical potential existing
for the system in both  the gauge-symmetric and the broken-symmetry phase.
At $T_c$, the chemical potential is continuous, $\mu(T_c-0)=\mu(T_c+0)$. One
should not confuse $\mu$ with $\mu_0$ which is just a Lagrange multiplier
guaranteeing the validity of the normalization condition (11), and keeping
the theory self-consistent and the system stable.

\section{Hartree-Fock-Bogoliubov Approximation}

We are now prepared to treat the Hamiltonian (18) with our modified 
HFB approximation. For this, we introduce some notations. The momentum 
distribution of particles
\be
\label{26}
n_\bk  \equiv \; \langle  a_\bk ^\dgr a_\bk  \rangle
\ee
will be referred to as the {\em normal average\/}, contrary to the {\em 
anomalous average\/}
\be
\label{27}
\sgm_\bk  \equiv \; \langle  a_\bk  a_{-\bk} \rangle  \; .
\ee
Summing these averages over momenta, one gets the density of uncondensed
particles
\be
\label{28}
\rho_1 \equiv \frac{1}{V} \; \sum_{\bk\neq 0} n_\bk
\ee
and the anomalous density
\be
\label{29}
\sgm_1 \equiv \frac{1}{V} \; \sum_{\bk\neq 0} \sgm_\bk  \; .
\ee

In the HFB approximation, the third-order term (22) is zero,
\be
\label{30}
~H^{(3)} ~= 0 \; ,
\ee
because of  (24). The fourth-order term (23), finally, becomes
\begin{eqnarray}
H^{(4)} &=& \sum_{\bk\neq 0} \rho_1 g \left ( a_\bk ^\dgr a_\bk  -\;
\frac{1}{2}\; n_\bk  \right )
\nonumber \\&+&
\label{31}
 \frac{1}{V} \; \sum_{\bk,\bp\neq 0} g \left [ n_{\bk+\bp} a_\bp^\dgr a_\bp +
\frac{1}{2}\left ( \sgm_{\bk+\bp} a_\bp^\dgr a_{-\bp }^\dgr +
\sgm_{\bk+\bp}^* a_{-\bp } a_\bp \right ) - \; \frac{1}{2} \left (
n_{\bk+\bp} n_\bp + \sgm_{\bk+\bp} \sgm_\bp^*\right ) \right ] \; .
\end{eqnarray}
Let us define  the shifted particle energies
\be
\label{32}
\om_\bk  \equiv \frac{k^2}{2m} + 2\rho g - \mu \; ,
\ee
where
\be
\label{33}
\rho \equiv \rho_0 + \rho_1
\ee
is the total particle density. We also  introduce the notation
\be
\label{34}
\Dlt \equiv (\rho_0 + \sgm_1) g \; .
\ee
Then the  Hamiltonian (18)
can be written in the  HFB approximation as
\be
\label{35}
H_{\rm HFB} = E_{\rm HFB} + \sum_{\bk\neq 0} \om_\bk  a_\bk ^\dgr a_\bk  +
\frac{1}{2} \; \sum_{\bk\neq 0} \Dlt \left ( a_\bk ^\dgr a_{-\bk} ^\dgr +
a_{-\bk}  a_\bk \right ) \; ,
\ee
where
\be
\label{36}
E_{\rm HFB} \equiv H^{(0)} - \; \frac{1}{2} \left ( 2\rho_1^2 + \sgm_1^2
\right ) g V \; .
\ee
Using this Hamiltonian in the grand-canonical  potential (12), we have
\be
\label{37}
\frac{\prt\Om}{\prt N_0} = (\rho +\rho_1 +\sgm_1) g - \ep \; ,
\ee
and  the second derivative is given by
\be
\label{38}
\frac{\prt^2\Om}{\prt N_0^2} = \frac{g}{V} \; .
\ee
Minimizing the grand-canonical  potential according to conditions (14), we obtain
\be
\label{39}
\ep = ( \rho +\rho_1 +\sgm_1) g  \; ,
\ee
with the stability condition $g>0$.

The Hamiltonian (35) is quadratic and can be diagonalized by means of the
Bogoliubov's canonical transformation
\be
\label{40}
a_\bk  = u_\bk  b_\bk  + v_{-\bk} ^* b_{-\bk} ^\dgr \; .
\ee
This brings the Hamiltonian (35) to the Bogoliubov form
\be
\label{41}
H_B = E_B + \sum_{\bk\neq 0} \ep_\bk  b_\bk ^\dgr b_\bk  \; ,
\ee
with the nonoperator energy term
\be
\label{42}
E_B \equiv E_{\rm HFB} + \frac{1}{2}\; \sum_{\bk\neq 0} (\ep_\bk  -\om_\bk )
\ee
and the Bogoliubov spectrum
\be
\label{43}
\ep_\bk  =\sqrt{\om_\bk ^2-\Dlt^2} \; .
\ee
By the Hugenholtz-Pines theorem \cite{11,12,13}, the spectrum
must be gapless:
\be
\label{44}
\lim_{k\ra 0} \ep_\bk  = 0 \; , \qquad \ep_\bk  \geq 0 \; .
\ee
 Inserting
 Eqs. (32) and (34) into (43), we obtain for the
chemical potential the equation
\be
\label{45}
\mu =(\rho +\rho_1 -\sgm_1) g \; .
\ee
It is easy to check that the same chemical potential (45) follows from the
Hugenholtz-Pines form $\mu=\Sigma_{11}(0,0)-\Sigma_{12}(0)$ by employing
Green function techniques. Both real-time or thermal Green functions can
be used, since they are just analytical continuations of each other \cite{34}.
Expression (45) is the standard form of the chemical potential in the HFB
approximation (see details in the review article [4]).

Comparing Eqs. (20), (39), and (45), we find that
the chemical potential of the condensate must satisfy
\be
\label{46}
\mu_0 = \ep -\mu = 2\sgm_1 g \; .
\ee
In the broken-symmetry phase, where the anomalous average $\sgm_1\neq 0$,
one has $\mu_0\neq 0$.
The value $\mu_0$ can be zero only for an ideal gas, when $g\ra 0$,
or in the Bogoliubov approximation [6,7], where the third- and fourth-order
Hamiltonian terms (22) and (23) are neglected. But in general, the Lagrange
multiplier $\mu_0$ is non-zero, thus making the theory self-consistent.

Using the chemical potential (45) in Eq. (32), we have
\be
\label{47}
\om_\bk  = \frac{{\bf k}^2}{2m} + \Dlt \; ,
\ee
and the spectrum (43) takes the form
\be
\label{48}
\ep_\bk  =\sqrt{(c{\bf k})^2 +\left ( \frac{{\bf k}^2}{2m}\right )^2} \; ,
\ee
with the sound velocity
\be
\label{49}
c \equiv \sqrt{\frac{\Dlt}{m}}.
\ee
In the long-wave limit $k\ra 0$, the quasiparticle
energy $\ep_\bk $ behaves like $ck$,  thus being
gapless, as it should.

For the normal average (26), we find
\be
\label{50}
n_\bk  =\left ( u_\bk ^2 + v_\bk ^2 \right ) f^{\rm b}_\bk  + v_\bk ^2 \; ,
\ee
while for the anomalous average (27), we get
\be
\label{51}
\sgm_\bk  = (1 +2f^{\rm b}_\bk ) u_\bk  v_\bk  \; ,
\ee
where the momentum distribution of bosonic quasiparticle  excitations
\be
\label{52}
f^{\rm b}_\bk  \equiv \; \langle b_\bk ^\dgr b_\bk \rangle  \; =
 \frac{1}{e^{\bt\ep_\bk}-1}
\ee
can be written in the form
\be
\label{53}
f^{\rm b}_\bk  = \frac{1}{2}\left(
{\rm coth} \frac{\ep_\bk }{2T}
-1 \right) .
\ee
The coefficient functions of the Bogoliubov transformation (40) satisfy
\be
\label{54}
u_\bk ^2 - v_\bk ^2 = 1 \; , \qquad
u_\bk  v_\bk  = -\; \frac{\Dlt}{2\ep_\bk } \; ,
\ee
and
\be
\label{55}
u_\bk ^2+ v_\bk ^2 = \frac{\sqrt{\ep_\bk ^2+\Dlt^2}}{\ep_\bk } =
\frac{\om_\bk }{\ep_\bk } \; ,
\ee
and read explicitly
\be
\label{56}
u_\bk ^2 =\frac{\sqrt{\ep_\bk ^2+\Dlt^2}+\ep_\bk }{2\ep_\bk } =
\frac{\om_\bk +\ep_\bk }{2\ep_\bk } ,
\ee
and
\be
\label{57}
v_\bk ^2 = \frac{\sqrt{\ep_\bk ^2+\Dlt^2}-\ep_\bk }{2\ep_\bk } =
\frac{\om_\bk -\ep_\bk }{2\ep_\bk } \; .
\ee
In this way, for the normal average (26), we obtain
\be
\label{58}
n_\bk  = \frac{\sqrt{\ep_\bk ^2+\Dlt^2}}{2\ep_\bk }\; {\rm coth}
\frac{\bt\ep_\bk }{2}- \frac{1}{2} \; ,
\ee
while the anomalous average (27) becomes
\be
\label{59}
\sgm_\bk  = -\; \frac{\Dlt}{2\ep_\bk }\; {\rm coth}
\frac{\bt\ep_\bk }{2} \; .
\ee

The anomalous average (59) is important as compared to the normal average
(58). For this purpose, consider these averages as functions of $\ep_\bk $
in the range
\be
\label{60}
0 \leq \ep_\bk .
\ee
At low momenta or energies, such that $\ep_\bk \ll\Dlt$ and $\ep_\bk \ll T$,
the asymptotic behavior of the normal average (58) is
\be
\label{61}
n_\bk  \simeq \frac{T\Dlt}{\ep_\bk ^2} + \frac{\Dlt}{12T} +
\frac{T}{2\Dlt}\; - \; \frac{1}{2} ,
\ee
whereas the anomalous average (59) is
\be
\sgm_\bk  \simeq  -\;\frac{T\Dlt}{\ep_\bk ^2}\; -\; \frac{\Dlt}{12T} +
\frac{\ep_\bk ^2\Dlt}{720T^3}\; .
\ee
From here it is evident that in the long-wave limit the anomalous average
is of the same order of magnitude as the normal one, $n_\bk \simeq |\sgm_\bk |$,
only their signs are opposite.

In the short-wave limit, when $\ep_\bk \gg\Dlt$ and $\ep_\bk \gg T$, the
asymptotic behavior of the normal average is
\be
\label{63}
n_\bk  \simeq \left ( \frac{\Dlt}{2\ep_\bk }\right )^2 - \left (
\frac{\Dlt}{2\ep_\bk }\right )^4 + e^{-\bt\ep_\bk } \; ,
\ee
whereas that of the anomalous average is
\be
\label{64}
\sgm_\bk  \simeq -\; \frac{\Dlt}{2\ep_\bk }\left ( 1 + 2 e^{-\bt\ep_\bk }
\right ) \; .
\ee
In this limit, the magnitude of the anomalous average becomes much
larger than the normal one.

Summarizing, we conclude that, in the above two limits, we have
\be\!\!\!\!\!\!\!\!\!\!\!\!\!\!\!\!
|\sgm_\bk | \simeq n_\bk  \qquad (k \ra 0) \; ,  ~~~~~~
\label{65}
|\sgm_\bk | \gg n_\bk  \qquad (k\ra \infty) \; .
\ee
Consequently, the anomalous average is always important.

In the large-volume limit, the sums (28) and (29) can be calculated as
momentum integrals, leading to the density of uncondensed particles (28):
\be
\label{66}
\rho_1 = \frac{1}{2} \; \int\frac{d^3k}{(2\pi)^3} \left [
\frac{\sqrt{\ep_\bk ^2+\Dlt^2}}{\ep_\bk }\; {\rm coth}
\frac{\bt\ep_\bk }{2} \; - \; 1 \right ] \;
 \; ,
\ee
while the anomalous density (29) becomes
\be
\label{67}
\sgm_1 = - \int \frac{d^3k}{(2\pi)^3} \, \frac{\Dlt}{2\ep_\bk }\; {\rm coth}
\frac{\bt\ep_\bk }{2} \; .
\ee
Taking into account relation (34) and using the notation
\be
\label{68}
\al \equiv \int\frac{d^3k}{(2\pi)^3} \; \frac{g}{2\ep_\bk } \;
{\rm coth}  \frac{\bt\ep_\bk }{2} \; ,
\ee
we obtain
\be
\label{69}
\Dlt = \frac{\rho_0g}{1+\al} \; ,
\ee
so that the anomalous density (67) can be represented in the
form
\be
\label{70}
\sgm_1 = -\; \frac{\rho_0\al}{1+\al} \; .
\ee

The quantity (68) is ultraviolet-divergent.
This
divergence is caused by the
modeling of the short-range interaction
in the dilute-gas approximation by a $ \delta $-function
potential (1).
There are two ways of removing this divergence. One may either use a
more realistic interaction potential $V(\br)$, whose Fourier
transform $V_\bk $ goes to zero  for $k\ra\infty$, for instance
Gaussian-type potentials \cite{35,36,37}. More efficiently,
with the same physical consequences
in the dilute limit,
one renormalizes the $ \delta $-function potential,
replacing it by the scattering matrix
obtained from
the Lippmann-Schwinger equation.
This simply renormalizes the coupling constant $g$ to the renormalized
$g_R$ defined by
\begin{equation}
\frac{1}{g_R}\equiv
\frac{1}{g}+\int \frac{d^3k}{(2\pi)^3}\frac{1}{\bk^2/2m-i0}.
\label{gR}
\end{equation}
Indeed, the relation to the scattering length in Eq. (2) is really only 
valid for the renormalized coupling constant $g_R$, which is usually not 
mentioned for brevity, until the ultraviolet divergences appear. This 
procedure is standard for eliminating divergences in calculating the 
ground-state energy \cite{2,3,4} as well as the anomalous averages 
\cite{38,39,40}. Keeping in mind such a correction, we may consider
the quantity (68) as finite.

It is easy to check that  condensation proceeds in a second-order phase
transition. When $\rho_0\ra 0$, $\rho_1\ra\rho$, we see that $\Dlt\ra 0$ and
$\ep_\bk \ra k^2/2m$. From Eq. (66) we find the condensation temperature
\be
T_c=\frac{2\pi}{m}\left[\frac{\rho}{\zeta(3/2)}\right]^{2/3},
\label{TC}
\ee
which coincides with that of the ideal gas, as it should be for a dilute gas
in mean-field approximation.

\section{Zero Temperature}

The above equations can be calculated explicitly in the zero-temperature
limit, where the density (66) becomes
\be
\label{71}
\rho_1 = \frac{1}{2} \; \int \frac{d^3k}{(2\pi)^3} \left (
\frac{\sqrt{\ep_\bk ^2 +\Dlt^2}}{\ep_\bk } \; - \; 1 \right ) \; .
\ee
This can also be represented as
\be
\label{72}
\rho_1 = \frac{1}{2}\; \int  \frac{d^3k}{(2\pi)^3} \left (
\frac{\om_\bk }{\ep_\bk } \; -\; 1 \right ) \;  .
\ee
Substituting Eqs. (47) and (48), we find
\be
\label{73}
\rho_1 = \frac{(mc)^3}{3\pi^2} \; .
\ee
For the fraction of condensed particles
\be
\label{74}
n_0 \equiv \frac{\rho_0}{\rho} =
 1 - n_1       =
 1 -   \frac{\rho_1}{\rho} ,
\ee
we obtain
\be
\label{75}
n_0 =  1 \; - \; \frac{(mca)^3}{3\pi^2} \qquad \left (
\rho a^3=1 \right ) \; ,
\ee
where $a$ is the mean interparticle distance.

The integral (68) becomes, after the renormalization of $g$ according to
(\ref{gR}),
\be
\label{76}
\al = \frac{g}{2}\; \int \left ( \frac{1}{\ep_\bk }\; - \;
\frac{2m}{k^2}\right ) \; \frac{d^3k}{(2\pi)^3} \; ,
\ee
where $g_R$ is sloppily replaced by $g$, as usual in such calculations. The 
value of $ \alpha $ is
\be
\label{77}
\al = -\; \frac{1}{\pi^2}\; m^2 c g \; ,
\ee
so that the anomalous average (70) becomes
\be
\label{78}
\sgm_1 = \frac{\rho_0 m^2 c\;g}{\pi^2-m^2 c\;g} \; .
\ee
This way of removing the ultraviolet divergences is quantitatively exact
for calculating quantities depending only on the scattering length $a_s$
in one-loop approximations [4]. For strong interactions, the shape of the
interaction potential becomes important [4,37--39]. In what follows, we shall
consider the anomalous average (80) for arbitrary $g$, but keep in mind that
at large $g$ this expression for $\sgm_1$ can be only qualitatively correct.
Being proportional to $\rho_0$, the anomalous density $\sgm_1$ tends to zero
for $\rho_0\ra 0$. However, in a dilute gas at low temperature, $\rho_0$ is
close to $\rho$ as follows from Eq. (77).

In further calculations, it is convenient to introduce
the {\em diluteness parameter\/}
\be
\label{79}
\dlt \equiv \rho^{1/3} a_s = \frac{a_s}{a}
\ee
and to introduce the reduced dimensionless sound velocity
\be
\label{80}
\hat c \equiv c a m \; .
\ee
With this notation,  the fraction of condensed particles is given by
\be
\label{81}
n_0 = 1 \; - \; \frac{\hat c^3}{3\pi^2} \; ,
\ee
and the fraction of uncondensed particles by
\be
\label{82}
n_1 = \frac{\hat c^3}{3\pi^2} \; .
\ee
Combining Eq. (79) with relation (2) for the renormalized $g$, we obtain
\be
\label{83}
\al= -\; \frac{4}{\pi}\; \hat c\dlt \; .
\ee
The anomalous average (80) is  then
\be
\label{84}
\sgm_1  = \frac{4\rho \hat c\dlt}{\pi-4\hat c\dlt}\; n_0 \; .
\ee

To define the above quantities (83)--(86) as functions of the diluteness 
parameter (81), we have to know the dependence on $\dlt$ of the reduced 
sound velocity (82). For this purpose, we use Eq. (69) in the form
\be
\label{85}
\Dlt = \frac{4\pi^2\dlt}{\pi-4\hat c\dlt} \left ( \frac{n_0}{ma^2}
\right ) \; ,
\ee
and recall that $\Dlt=mc^2$ according to  Eq. (49). Thus Eq.~(87) becomes
\be
\label{86}
8\dlt \hat c^3 - 3\pi \hat c^2 + 12\pi^2 \dlt = 0 \; ,
\ee
which fixes the reduced sound velocity as a function $\hat c(\dlt)$ of the 
diluteness parameter. There exists  a positive solution of Eq. (88) for 
$\hat c$ in the $ \delta $-interval $0\leq\dlt\leq\dlt_c$, limited by the 
critical value
\be
\label{87}
\dlt_c \equiv \frac{1}{4} \left ( \frac{\pi}{3}\right )^{1/3} =
0.253873 \; .
\ee

Remarkably, there exists an interesting relation between the anomalous 
average (86) and the density of uncondensed particles $\rho_1$. We take the 
identity following from  $\Dlt=mc^2$:
\be
\label{88}
\Dlt ma^2 =(mac)^2 = \hat c^2 \; ,
\ee
and substitute on the left-hand side $\Dlt$ from Eq. (87) to obtain
\be
\label{89}
\hat c^2 = \frac{4\pi^2\dlt n_0}{\pi-4\hat c\dlt} \; .
\ee
Inverting this with respect to $\dlt$ yields
\be
\label{90}
\dlt = \frac{\pi \hat c^2}{4(\pi^2 n_0 + \hat c^3)} \; .
\ee
Substituting Eq. (92) into Eq. (86), we obtain
\be
\label{91}
\sgm_1 = 3\rho_1 \; .
\ee
That is, the anomalous average is three times larger than the normal
one. This emphasizes once more that at low temperatures the anomalous
averages are important.

At asymptotically small diluteness parameter $\dlt\ll 1$, the reduced 
sound velocity (82) in  Eq. (88) behaves as
\be
\label{92}
\hat c \simeq 2\sqrt{\pi}\; \dlt^{1/2} + \frac{16}{3}\; \dlt^2 +
\frac{320}{9\sqrt{\pi}}\; \dlt^{7/2} \qquad (\dlt\ll 1) \; ,
\ee
while the condensate fraction (83) has the expansion
\be
\label{93}
n_0 \simeq 1 \; - \; \frac{8}{3\sqrt{\pi}} \; \dlt^{3/2} - \;
\frac{64}{3\pi} \; \dlt^3  \; .
\ee
Keeping only the first term in the expansion (94), we obtain the Bogoliubov
sound velocity $c_B=\sqrt{4\pi\rho a_s}/m$. Retaining on the right-hand side 
of Eq. (95) the first two terms, we obtain the Bogoliubov depletion formula 
[6--9]. In the Sec. V, we show that the ground-state energy in the limit 
$\dlt\ra 0$ also gives the known Bogoliubov expression. Thus, in the limit 
of the small diluteness parameter $\dlt$, our equations have the correct 
asymptotic behavior, reproducing the results of the Bogoliubov approximation.

When the diluteness parameter approaches the critical value (89), the sound 
velocity $c=\hat c/am$ has the expansion
\be
\label{94}
\hat c \simeq \hat c_c - 2\sqrt{3\pi} (\dlt_c - \dlt)^{1/2} + 8\left (
\frac{\pi}{3}\right )^{1/3} (\dlt_c -\dlt) \qquad
( \dlt \ra \dlt_c - 0 ) \; ,
\ee
with
\be
\label{95}
\hat c_c \equiv \left ( 3\pi^2\right )^{1/3} = 3.093 668 \; .
\ee
And the condensate fraction is given by
\be
\label{96}
n_0 \simeq 6 \left ( \frac{3}{\pi}\right )^{1/6} (\dlt_c -
\dlt)^{1/2} - 20 \left ( \frac{3}{\pi}\right )^{1/3} (\dlt_c - \dlt)
\qquad (\dlt\ra \dlt_c - 0 ) \; .
\ee
At the critical depletion (89), we have
\be
\label{97}
c = \frac{(3\pi^2)^{1/3}}{am} \; , \qquad n_0 = 0 \qquad
( \dlt = \dlt_c) \; .
\ee

The disappearance of the condensate fraction for $\dlt\geq \dlt_c$ at
zero temperature is a signal for a {\em quantum phase transition\/}. Here
this transition occurs as a function of the diluteness parameter $\dlt$.
To display the critical behavior of physical properties, it is convenient to 
introduce the relative distance variable from the quantum critical point
\be
\label{98}
\tau \equiv \frac{\dlt-\dlt_c}{\dlt_c} \; .
\ee
Then the reduced sound velocity behaves in the vicinity of the critical
point as
\be
\label{99}
\frac{\hat c-\hat c_c}{\hat c_c} \simeq -\tau^{1/2} + \frac{2}{3}\; \tau 
\qquad (\tau \ll 1) \; .
\ee
For the condensate fraction we obtain
\be
\label{100}
n_0 \simeq 3\tau^{1/2} -  5\tau \qquad (\tau \ll 1) \; ,
\ee
implying that  the critical exponent $  \beta  $ for the order parameter 
$\Psi$ is $1/4$.

The overall behavior of the reduced sound velocity $\hat c=\hat c(\dlt)$
and of the condensate fraction $n_0=n_0(\dlt)$ as functions of the
diluteness parameter $\dlt$ are shown in Figs. 1 and 2, respectively.
The function $\hat c(\dlt)$ in Fig. 1 is calculated from Eq. (88).
Substituting this $\hat c(\dlt)$ into Eq. (83), we get the condensate 
fraction $n_0(\dlt)$ plotted in Fig. 2.

\section{Thermodynamic Consistency}

As a final important point we convince ourselves that our self-consistent 
bosonic HFB approximation involving the new Lagrange multiplier $\mu_0$
is consistent with the thermodynamic formalism, in contrast to previous 
attempts to generalize the HFB approximation to condensed Bose systems.
For the discussion of the inconsistencies we mention once more the work of
Hohenberg and Martin [21] and the subsequent papers [17--19,30--32,43].
The most recent and very clear analysis of the problem was given in the 
review article by Andersen [4].

The grand-canonical potential $ \Omega $ in Eq.~(12) and the associated
Hamiltonian $H$ in Eq. (13) make it possible to find all thermodynamic 
properties of the system. The free energy (9) is connected with the grand 
potential (12) through the relation
\be
\label{101}
\Om = F - \mu_0 N_0  - \mu N \; .
\ee
The Lagrange multipliers $\mu_0$ and $\mu$ should not be confused with the
standard chemical potential defined for the system without condensate.
In the presence of a condensate, its role is played by what we may call
{\em effective system chemical potential\/}, denoting it by $\tilde \mu$.
In the condensed phase, it is given by
\be
\label{102}
\tilde\mu \equiv \mu_0 n_0 +\mu\; ,
\ee
where $n_0\equiv N_0/N$. With this definition, the grand-canonical  
potential (103) satisfies the usual thermodynamic relation
\be
\label{103}
\Om =  F - \tilde\mu N \;.
\ee
Its differential satisfies
\be
\label{104}
d\Om = - S dT - P dV - N d\tilde\mu  \; ,
\ee
where $S$ is entropy and $P$ pressure. For the free energy $F$, this implies
\be
\label{105}
dF = - S dT - P dV + \tilde\mu dN \; .
\ee
Thus the system chemical potential $\tilde \mu$ is given by the derivative
\be
\label{106}
\tilde\mu  = \left ( \frac{\prt F}{\prt N}\right )_{TV} \; .
\ee
From these expressions, we may calculate all thermodynamic properties of 
the condensed Bose gas. Thus our self-consistent HFB approximation obeys 
the standard thermodynamic formalism.

As an illustration, let us verify the consistency of the above two definitions 
(104) and (108) for the system chemical potential. To be explicit, consider 
a dilute Bose gas at zero temperature, representing the universal terms [4], 
which are asymptotically exact for all short-range interaction potentials 
with scattering length $a_s$.

At zero temperature, the free energy coincides with the internal ground-state 
energy, $F=E$. Then Eq. (108) reduces to
\be
\label{107}
\tilde\mu  =\left ( \frac{\prt E}{\prt N}\right )_V \qquad (T=0) \; .
\ee
The internal energy is, by definition,
\be
\label{108}
{E} = {\langle H\rangle} + \tilde\mu N\; ,
\ee
where $H$ is the grand-canonical  Hamiltonian (13). The average 
$\langle H\rangle$ is given in the HFB approximation by
\be
\label{109}
\langle H \rangle\; = \; E_B  = E_{\rm HFB} +
V \int \frac{d^3k}{(2\pi)^3} (\ep_\bk  - \xi_\bk ) \; ,
\ee
where, according to Eqs. (42) and (19),
\be
\label{111}
{E_{\rm HFB}} = {H^{(0)}} \; - \; \frac{gV}{2}\,
\left ( 2\rho_1^2 + \sgm_1^2 \right ) \; =
N_0 \left ( \frac{1}{2}\; \rho_0 g - \mu_0 - \mu \right )
- \; \frac{gV}{2}\, \left ( 2\rho_1^2 + \sgm_1^2 \right ) \; .
\ee
For the explicit calculation of the integral in (111) in the dilute gas, 
we perform the usual subtraction implied by the renormalization of $g$ in 
Eq.~(\ref{gR}). The subtracted integral becomes
\be
\label{113}
\frac{1}{2} \; \int \frac{d^3k}{(2\pi)^3} \left [ {\ep_\bk
-\xi_\bk } + \frac{m^3c^4}{k^2}\right ] = \frac{8m^4c^5}{15\pi^2}\;  .
\ee
Employing the results of Sec. IV, we find for the internal energy
(110), which at $T=0$ is the ground-state energy,
\be
\label{114}
{E} \simeq \frac{g N^2}{2V} \left ( 1 +
\frac{128}{15\sqrt{\pi}} \; \dlt^{3/2} \right ) \; .
\ee
This is the expression derived by Lee et al. [44,45] for a hard-sphere
potential. It is universal in the sense that it applies to any short-range 
potential with scattering length $a_s$ [4]. Differentiating (114) with 
respect to $N$ we obtain, according to (109) and using the fact that $F=E$ 
at $T=0$, the effective chemical potential
\be
\label{115}
\tilde\mu \simeq \rho g \left ( 1 + \frac{32}{3\sqrt{\pi}}\;
\dlt^{3/2}\right ) \; ,
\ee
where we have inserted the derivative $\prt\dlt/\prt N =\dlt/3N$. The same
result is obtained from (104), showing
the self-consistency of our HFB approximation.

It is important to emphasize that the self-consistency in our approach
has been achieved by accurately taking into account all normalization
conditions, which required the introduction of the additional Lagrange
multiplier $\mu_0$. It is due to the latter that our HFB approximation
is completely self-consistent and displays no gap in the spectrum.
Without $\mu_0$, we would plunge back to the known problem of the
standard HFB approximation, which is not self-consistent and possesses
an unphysical gap in the spectrum [17--19,30--32,43].

In conclusion, we have presented a new bosonic self-consistent
Hartree-Fock-Bogoliubov approximation. It is derived from a variational
principle and preserves all symmetries of the Hamiltonian. At the same 
time, it is gapless in the condensed phase, thus solving an old outstanding
problem of Bose systems. We did not invoked any unjustified tricks as
in previous attempts with the same goal, such as omitting anomalous 
averages, and avoided any mismatch of approximations by adding additional 
phenomenological terms to remove the gap. This became possible by accurately 
taking into account two normalization conditions. Thus our  HFB 
approximation is conserving, gapless, and self-consistent.

\vskip 5mm

{\bf Acknowledgment}. We thank E.P. Yukalova for numerical calculations
and useful advice. One of the authors (V.I.Y.) is grateful for financial
support of the German Research Foundation and to the Physics Department of
the Freie Universit\"at Berlin for its hospitality.

\newpage

\begin{figure}[h]
\unitlength1mm
\begin{picture}(10.64,79.645)
\put(0,0){\psfig{file=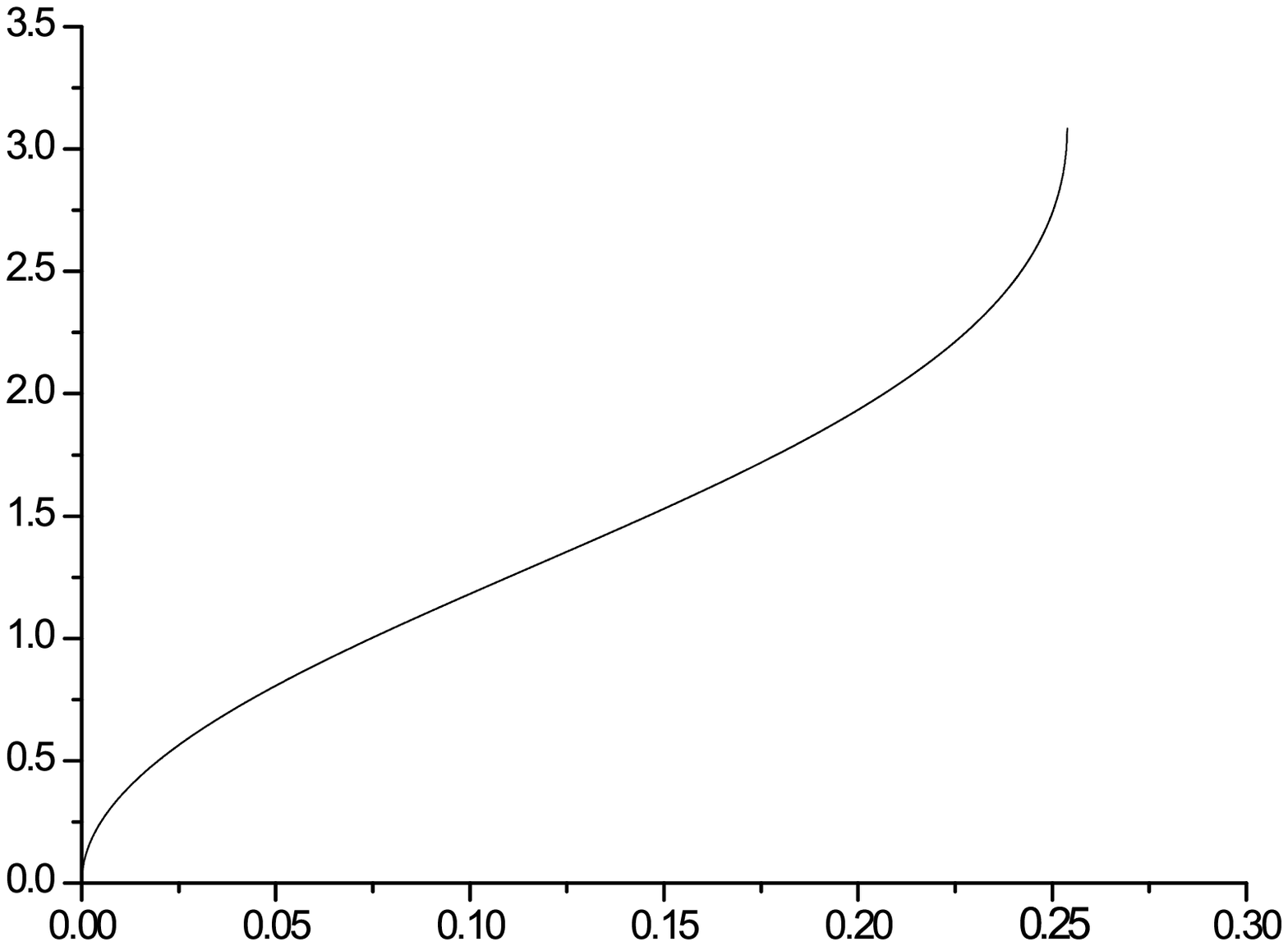,width=11cm}}
\put(80,8){{\footnotesize$ \delta _c=0.254$}}
\put(82.3,11){\vector(0,1){2}}
\put(15,71){{\footnotesize$\hat c_c=3.0937$}}
\put(15,71.7){\vector(-1,0){2}}
\put(44.2,49){{$\hat c( \delta )$}}
\end{picture}
\caption{The reduced sound velocity $\hat c=\hat c(\dlt)$ as a function
of the diluteness parameter $\dlt$, which is given by the solution of
Eq. (88).}
\label{Fig.1}
\end{figure}

\begin{figure}[h]
\unitlength1mm
\begin{picture}(10.64,79.645)
\put(1,2){\psfig{file=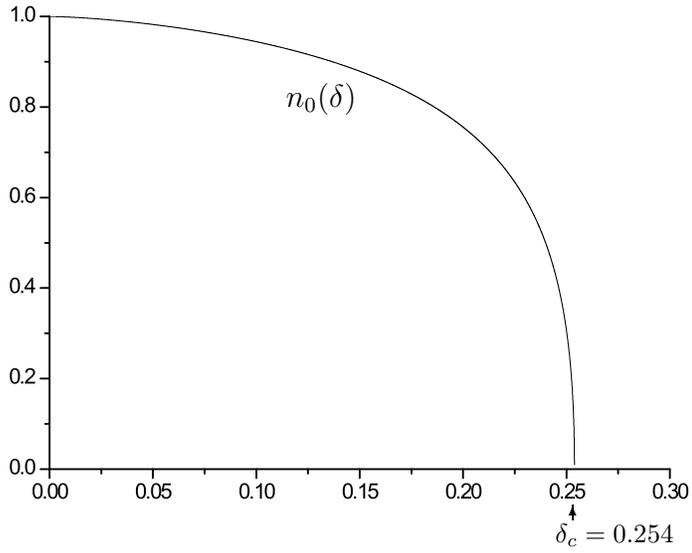,width=11cm}}
\put(80,8){{\footnotesize$ \delta _c=0.254$}}
\put(82.3,11){\vector(0,1){2}}
\put(44.2,66){{$n_0( \delta )$}}
\end{picture}
\caption{The condensate fraction $n_0=n_0(\dlt)$ as a function of
$\dlt$, obtained from Eq. (83).}
\label{Fig.2}
\end{figure}

\newpage

{\large{\bf Figure Captions}}

{\parindent=0pt
\vskip 2cm

{\bf Fig. 1}. The reduced sound velocity $\hat c=\hat c(\dlt)$ as a function
of the diluteness parameter $\dlt$, which is given by the solution of
Eq. (88).

\vskip 2cm

{\bf Fig. 2}. The condensate fraction $n_0=n_0(\dlt)$ as a function of
$\dlt$, obtained from Eq. (83).
}

\end{document}